%% file: main.tex
\documentclass[sigconf,screen=true]{acmart}

\AtBeginDocument{%
  }

\makeatother

\copyrightyear{2024} 
\acmYear{2024} 
\setcopyright{rightsretained} 
\acmConference[WSDM '24]{Proceedings of the 17th ACM International Conference on Web Search and Data Mining}{March 4--8, 2024}{Merida, Mexico}
\acmBooktitle{Proceedings of the 17th ACM International Conference on Web Search and Data Mining (WSDM '24), March 4--8, 2024, Merida, Mexico}\acmDOI{10.1145/3616855.3635770}
\acmISBN{979-8-4007-0371-3/24/03}

\makeatletter
\gdef\@copyrightpermission{
 \begin{minipage}{0.3\columnwidth}
  \href{https://creativecommons.org/licenses/by/4.0/}{\includegraphics[width=0.90\textwidth]{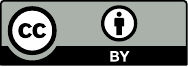}}
 \end{minipage}\hfill
 \begin{minipage}{0.7\columnwidth}
  \href{https://creativecommons.org/licenses/by/4.0/}{This work is licensed under a Creative Commons Attribution International 4.0 License.}
 \end{minipage}
 \vspace{5pt}
}
\makeatother

\author{Xiaojie Sun}
\affiliation{
	\institution{CAS Key Lab of Network Data Science and Technology, ICT, CAS}
	\institution{University of Chinese Academy of Sciences}
	\city{Beijing}
	\country{China}
}
\email{sunxiaojie21s@ict.ac.cn}

\author{Keping Bi}
\affiliation{
	\institution{CAS Key Lab of Network Data Science and Technology, ICT, CAS}
	\institution{University of Chinese Academy of Sciences}
	\city{Beijing}
	\country{China}
}
\email{bikeping@ict.ac.cn}

\author{Jiafeng Guo}
\authornote{Jiafeng Guo is the corresponding author.}
\affiliation{
	\institution{CAS Key Lab of Network Data Science and Technology, ICT, CAS}
	\institution{University of Chinese Academy of Sciences}
	\city{Beijing}
	\country{China}
}
\email{guojiafeng@ict.ac.cn}

\author{Sihui Yang}
\affiliation{
	\institution{CAS Key Lab of Network Data Science and Technology, ICT, CAS}
 \institution{University of Chinese Academy of Sciences}
 \city{Beijing}
 \country{China}
}
\email{yangsihui22s@ict.ac.cn}

\author{Qishen Zhang}
\author{Zhongyi Liu}
\author{Guannan Zhang}
\affiliation{%
  \institution{Ant Group}
  \city{Beijing}
  \country{China}}
\email{{qishen.zqs, zhongyi.lzy, zgn138592}@alibaba-inc.com}

\author{Xueqi Cheng}
\affiliation{
	\institution{CAS Key Lab of Network Data Science and Technology, ICT, CAS}
 \institution{University of Chinese Academy of Sciences}
 \city{Beijing}
 \country{China}
}
\email{cxq@ict.ac.cn}

\usepackage{acronym}

\usepackage{tablefootnote}

\usepackage{bm}
\usepackage{caption}
\usepackage{graphicx}
\usepackage{subfigure}
\usepackage{amsmath}
\usepackage{booktabs}
\usepackage{threeparttable}
\usepackage{multirow}
\usepackage{enumitem}
\usepackage{xcolor}
\usepackage{mathtools}
\usepackage{marvosym}
\usepackage[ruled]{algorithm2e}
\usepackage{tablefootnote}
\usepackage{soul, color, xcolor}

\newcommand{\paratitle}[1]{\vspace{1.5ex}\noindent \textbf{#1}}
\newcommand{\baby}{\textsc{MURAL}\xspace}
\newcommand{\babyw}{\textsc{MUR}\xspace}
\newcommand{\alipay}{{Alipay}\xspace}

\newcommand{\eg}{\emph{e.g.,}\xspace}

\newcommand{\eat}[1]{}
\begin{document}

\title{A Multi-Granularity-Aware Aspect Learning Model for Multi-Aspect Dense Retrieval}


\renewcommand{\shortauthors}{Xiaojie Sun et al.}


\begin{abstract}
Dense retrieval methods have been mostly focused on unstructured text and less attention has been drawn to structured data with various aspects, e.g., products with aspects such as category and brand. Recent work has proposed two approaches to incorporate the aspect information into item representations for effective retrieval by predicting the values associated with the item aspects. Despite their efficacy, they treat the values as isolated classes (e.g., ``Smart Homes'', ``Home, Garden \& Tools'', and ``Beauty \& Health'') and ignore their fine-grained semantic relation. 
Furthermore, they either enforce the learning of aspects into the \texttt{CLS} token, which could confuse it from its designated use for representing the entire content semantics, or learn extra aspect embeddings only with the value prediction objective, which could be insufficient especially when there are no annotated values for an item aspect. 

Aware of these limitations, we propose a MUlti-granulaRity-aware Aspect Learning model (\baby) for multi-aspect dense retrieval. It leverages aspect information across various granularities to capture both coarse and fine-grained semantic relations between values. Moreover, \baby incorporates separate aspect embeddings as input to transformer encoders so that the masked language model objective can assist implicit aspect learning even without aspect-value annotations. Extensive experiments on two real-world datasets of products and mini-programs show that \baby outperforms state-of-the-art baselines significantly. Code will be available at the URL\footnote{https://github.com/sunxiaojie99/MURAL}.

\end{abstract}

\ccsdesc[500]{Information systems~Information retrieval}

\keywords{Dense Retrieval, Multi-Aspect, Pre-training}

\maketitle

\vspace*{-2mm}
\section{Introduction}
\label{intro}


\begin{figure}[h]
\setlength{\belowcaptionskip}{-0.3cm}
\setlength{\abovecaptionskip}{0pt}
\includegraphics[scale=0.28]{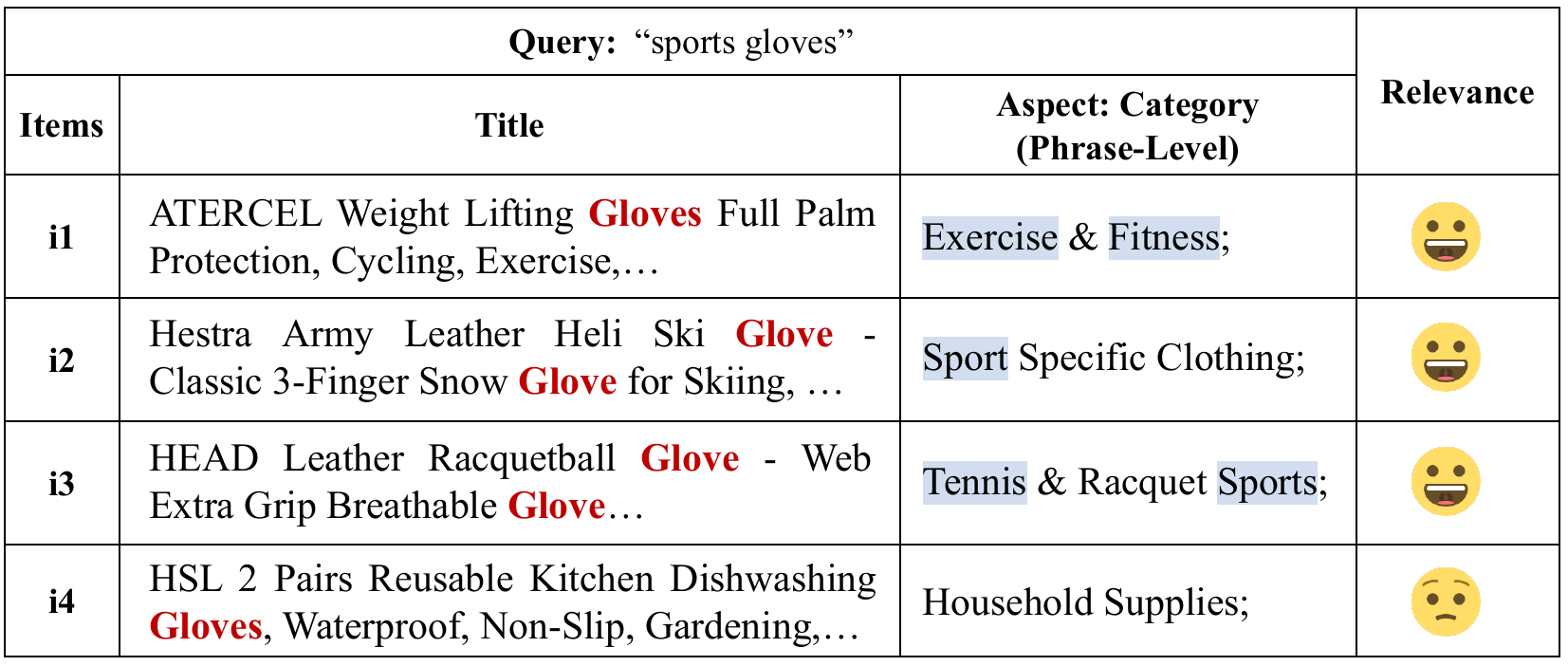}
\caption{
An example of a query and its candidate items. 
}
\label{fig:example}
\end{figure}

In recent years, dense retrieval methods have been extensively studied in both Information Retrieval (IR) and Natural Language Processing (NLP) communities \cite{retrieval-review}. On the shoulders of pre-trained language models (PLMs), they have achieved compelling performance. However, they are mostly studied for unstructured data and have not investigated how to effectively leverage the aspect information of structured data, such as category for products and affiliation for people. For example, in Figure \ref{fig:example}, the query ``sports gloves'' targets gloves for sports use so kitchen gloves should be avoided. It is obvious that the category of the four items could help to differentiate various types of gloves and improve retrieval performance. Unfortunately, it remains largely unexplored to effectively leverage such aspect information in dense retrieval. 

Recently, \citet{madr} has proposed two effective models for multi-aspect dense retrieval, i.e., MTBERT and MADRAL. These methods follow a typical paradigm of learning aspect embeddings with an auxiliary objective of predicting their associated values \cite{rerank-explain,rerank-explain2}. A concrete example is that the embedding of aspect ``category'' for i4 in Figure \ref{fig:example} will be learned by predicting its value, i.e., ``Household Supplies''. Although effective, they consider the values of an aspect as isolated classes and neglect the potential correlation between various values, which could result in sub-optimal performance. Taking the items in Figure \ref{fig:example} for instance, although they fall into four separate categories, the first three are relevant to the user query ``sports gloves'' while the last is not. The auxiliary objective of predicting their categorical IDs treats each category equally and may not capture their fine-grained relations. 

Noticing this issue, we propose to leverage the aspect information at even finer granularities, such as the word and token levels, in addition to the previously considered phrase-level granularity. Then, for the items in Figure \ref{fig:example}, when we break their category phrases into small pieces, the relation between the first three will be clearer since they all have sports-related descriptions such as exercise, sport, tennis, etc. Moreover, from a linguistic perspective, coarser granularities such as sentences and phrases convey more specific information while finer units usually carry more general information \cite{granu}. Since different granularities could express various levels of intent, we incorporate multiple granularities of aspect annotation prediction to assist query/item representation learning. 

Our model is named \baby, short for a MUlti-granulaRity-aware Aspect Learning model. It incorporates separate aspect embeddings before the content tokens and after \texttt{CLS} as inputs to the transformer layers (shown in Figure \ref{fig:model}). Then, on the top layer, the aspect embeddings are supervised with value predictions at various levels of granularities (e.g., phrase, word, and token). \baby has several advantages over state-of-the-art methods, i.e., MTBERT and MADRAL (See Figure \ref{fig:al_pic}): 
First, in contrast to MTBERT which mixes the information from item aspects and the overall content semantics in \texttt{CLS}, \baby represents the two types of information separately and allows for more interactions between them with a gating mechanism.
Second, in contrast to MADRAL which only learns the aspect embeddings with the value prediction objective during pre-training, \baby also guides the aspect embeddings to learn from the masked language model loss. This could assist implicit aspect learning even when there are no annotated values for an item aspect. 
%
Last and most importantly, by incorporating the aspect information across various granularities, \baby could capture the semantic relations between the aspect values at different levels, contributing more to the retrieval performance. 

We conduct extensive experiments on two real-world search datasets with rich aspect information.
Experimental results show that our method outperforms competitive baselines significantly on both datasets. It is remarkable that our model achieves compelling performance even without the supervision of aspect annotations, which means that useful implicit representations can be learned by \baby even when the aspect information is not used. Ablation studies on different granularities show that each granularity can contribute to the multi-aspect retrieval performance and combining them all lead to much better results. 

\input{related}

\input{method}

\input{exp-settings}

\input{exp-results}

\vspace*{-2mm}
\section{CONCLUSIONS}
In this paper, we propose a multi-granularity-aware aspect learning model that enhances the utilization of additional aspect information in structured data. Unlike previous methods that disregard the semantic relationship among different aspect values, our approach incorporates multiple granularities of aspect values to facilitate query/item representation learning. 
By effectively capturing the semantics of queries/items from implicit views, our model achieves compelling performance even without the supervision of aspect annotations.
Empirical results on two real-world datasets demonstrate the superiority of \baby. 

\vspace*{-2mm}
\begin{acks}
This work was funded by the National Natural Science Foundation of China (NSFC) under Grants No. 62302486, the Innovation Project of ICT CAS under Grants No. E361140, the CAS Special Research Assistant Funding Project, the Lenovo-CAS Joint Lab Youth Scientist Project, and the project under Grants No. JCKY2022130C039. This work was also supported by Ant Group through Ant Innovative Research Program.
\end{acks}

\bibliographystyle{ACM-Reference-Format}
\bibliography{paper}

\end{document}

%% file: related.tex
\vspace*{-3mm}
\section{RELATED WORK}




\noindent
\textbf{Dense Retrieval.}
Dense retrieval models typically use a bi-encoder structure for independent query and item encoding, with relevance measured through a simple similarity function (such as dot product). 
\citet{DPR} initializes the encoder with BERT and combines it with in-batch negatives, 
achieving better performance than early models. 
After that, researchers began to explore various fine-tuning techniques to train a better dense retriever, including hard negative mining \cite{ance,rocketqa}, knowledge distillation \cite{distil-cross}, and multi-vector representation \cite{me-bert, col-bert, mvr}. 
For example, \citet{ance} proposed to dynamically mine hard negatives during training by periodically refreshing the index. 
\citet{me-bert} captures information of items from different perspectives by using the first k document token embeddings as the item representation. 
Based on this, \citet{mvr} added k special tokens before the item input to obtain the multi-vector representation. 
These multi-vector methods aim to extract multiple underlying semantic information from the item. In contrast, our method explicitly considers explicit multi-aspect information modeling. Additionally, our method outputs only a single representation vector for each item, saving space and time for indexing items.

Recently, \citet{madr} introduced two methods for incorporating explicit aspect information into a single representation vector. The first method employs \texttt{CLS} embeddings to simultaneously perform aspect classification tasks for multiple aspects. The second method adds an attention network to the PLM, enabling it to separately model multiple aspects, followed by aspect fusion. 
Their differences with our method will be introduced in Section \ref{sec:method}.

\noindent
\textbf{Multi-Field Retrieval.}
The effective utilization of multi-field information (\eg title, keyword, description) in documents has been studied for long. 
Before PLM appears, many neural ranking models were proposed to effectively leverage item structure \cite{field-Balaneshinkordan,field-liu,field-Zamani}.
For example, \citet{field-Zamani} aggregated field-level representations to obtain item representations and employed a matching network for final relevance score prediction. 
In the PLM era, research has continuously focused on the utilization of multi-field information \cite{field-shan2023beyond, sun2023cikm}.
For example, 
\citet{field-shan2023beyond} proposed the field-level local matching loss, calculated based on the query and each document field representation.
\citet{sun2023cikm} treated aspect as text and proposed an effective pre-training method to capture the bi-directional interactions between aspect and content texts.
The difference between multi-aspect and multi-field is that fields contain an infinite textual value space, usually composed of variable-length unstructured text. 
Conversely, an aspect has a defined set of finite values, acting as "labels" for structured items. 
Given this, they face different core challenges, and effectively utilizing multiple aspects' information is a valuable research direction. 

\noindent
\textbf{Pre-trained Bi-encoder.}
Researchers have explored pre-training models for retrieval with the bi-encoder architecture\cite{pre-LeeCT19,pre-seed,condenser,costa, co-condenser, pre-retromae}.
For example, 
\citet{condenser} added extra head layers atop the Transformer, with shortcut connections between early outputs and the head, enhancing the \texttt{CLS} embedding of the encoder.
\citet{pre-seed} pre-trained an auto-encoder with a weak decoder for document representation learning.
Differing from these pre-training methods targeted at unstructured data, we investigate how to infuse explicit aspect information into the encoder representation during pre-training. 
In the future, we will explore how to integrate our approach with existing research.

%% file: method.tex

\vspace*{-2mm}
\section{Preliminaries}
\label{sec:Preliminaries}

\vspace*{-2mm}
\paratitle{Dual Encoding.} 
The standard PLMs, \eg BERT \cite{bert}, take a token sequence $X=(x_1, ..., x_n)$ as input, and generate contextualized representations as:
\begin{equation}
\setlength{\belowcaptionskip}{0cm}
\setlength{\abovecaptionskip}{0pt}
\small
\textbf{h}(x_{0}),\textbf{h}(x_{1})... \textbf{h}(x_{n})= \Phi_{enc}(X), \  \textbf{h}(x_i) \in \mathbb{R}^H,
\end{equation}
where $H$ denotes the hidden size, and $x_0=[CLS]$ is a special token added to the beginning. 
The representation $\textbf{h}(x_0)$ is commonly used as the final representation for the input $X$.
In dense retrieval, the bi-encoder architecture is widely adopted, where the query $Q$ and item $I$ are separately encoded using the PLM to obtain their respective representation vectors\cite{dual-encoder}. 
Then, a simple scoring function is used to calculate the similarity between these two vectors.
\renewcommand{\arraystretch}{0.5}
\begin{table}
\setlength{\abovecaptionskip}{0pt}
  \small
  \caption{A summary of main notations used in this paper. 
  }
  \label{tab:notations}
  \begin{tabular}{p{3cm}p{4.8cm}}
    \toprule
    \textbf{Notation} &\textbf{Meaning}\\
    \midrule
    $X=(x_1,x_2,...,x_n)$ & The input token sequence of query/item.\\
    \midrule
    $\textbf{h}_{X}$ & The final representation for the input X.\\
    \midrule
    \parbox{3cm}{$A = \{a_i\}$, $i=1, ..., |A|$} & \parbox{4.8cm}{A set of aspects, \eg \{\textit{brand}, \textit{color}, \textit{category}\}}\\
    \midrule
    \parbox{3cm}{$G = \{g_i\}, i=1, ..., |G|$} & \parbox{4.7cm}{The set of language granularities, \eg $\{phrase, word, token\}$}\\
    \midrule
    \parbox{3cm}{$V_{a_i}^{g_j}$ or $V_a^g$, \\  $i=1, ..., |A|$, $j=1, ..., |G|$} & \parbox{4cm}{The aspect value vocabulary of aspect $a_i$ at granularity level $g_j$, \eg  $V_{category}^{word}=\{shoes, clothing,...\}$} \\
    \midrule
    \parbox{3cm}{$\textbf{h}_{a_i}^{g_j}$ or $\textbf{h}_a^g$, \\ $i=1, ..., |A|$, $j=1, ..., |G|$} & \parbox{4.7cm}{The aspect embedding for query's or item's aspect $a$ at granularity $g$.}\\
    \midrule
    \parbox{3cm}{$E_{a_i}^{g_j}$ or $E_a^g$, \\ $i=1, ..., |A|,j=1, ..., |G|$} 
    & \parbox{4.7cm}{The aspect value embedding table for aspect $a$ at granularity $g$.}\\
    \midrule
    \parbox{3cm}{$\mathcal{A}_{a_i}^{g_j}$ or $\mathcal{A}_a^g$, \\ $i=1, ..., |A|,j=1, ..., |G|$} & \parbox{4.7cm}{The set of aspect value annotations for query's or item's aspect $a$ at granularity $g$, $\mathcal{A}_a^g \subset V_a^g$} \\
  \bottomrule
\end{tabular}
\end{table}

\begin{figure}[t]
\setlength{\belowcaptionskip}{0cm}
\setlength{\abovecaptionskip}{0pt}
\includegraphics[scale=0.3]{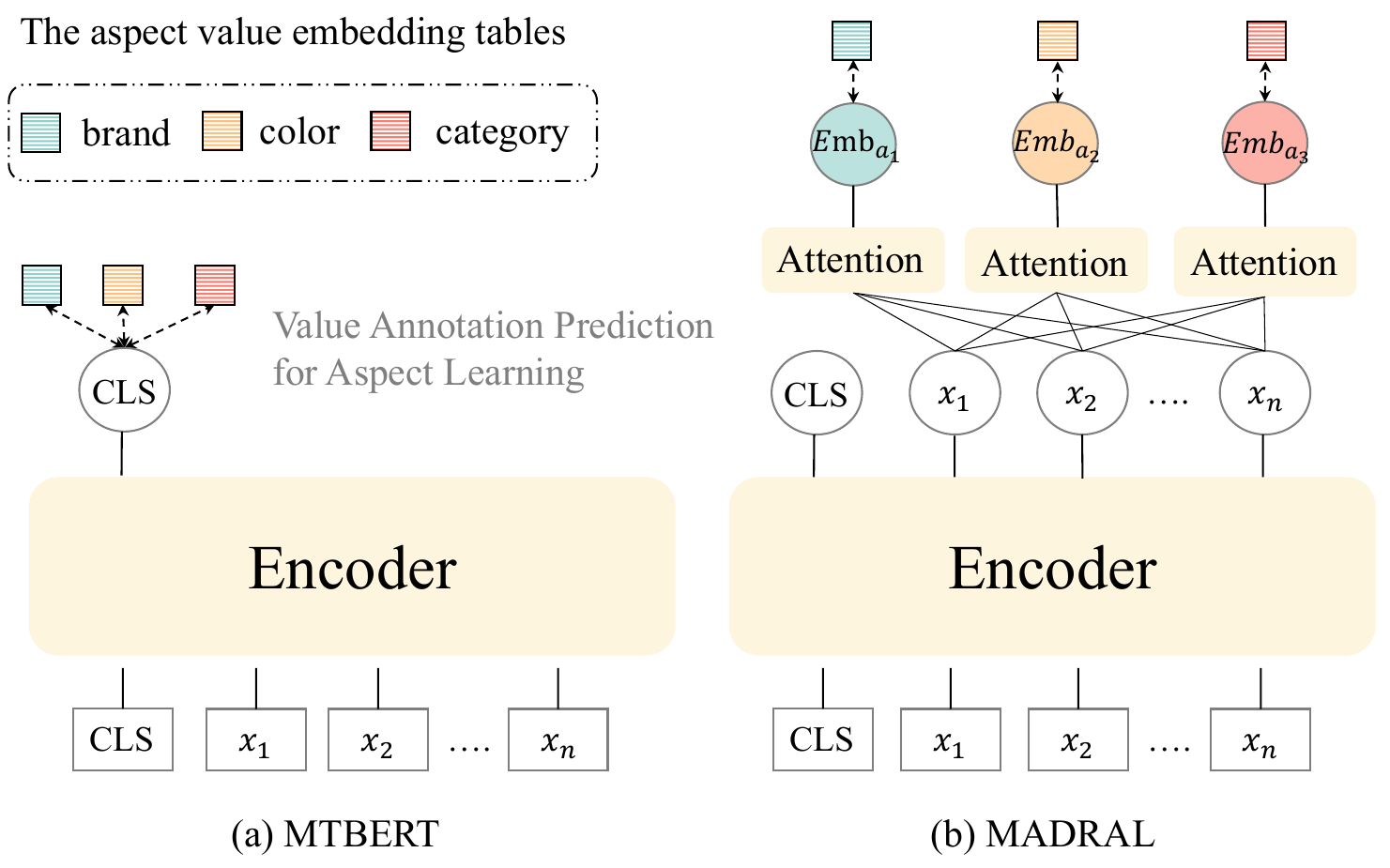}
\caption{
SOTA multi-aspect dense retrieval models.
}
\label{fig:al_pic}
\end{figure}

\vspace*{-2mm}
\paratitle{Aspect Learning.} 
In dense retrieval, aspect learning involves using aspect information to enhance retrieval performance when queries or items are associated with varying aspects (e.g., \textit{brand}, \textit{color}, \textit{category} in product search).
In addition to the content text $X=(x_1, ..., x_n)$ (e.g., query, item title), a query or item can be associated with multiple aspects, and we denote the set of these aspects as $A=\{a_i\}_{i=1}^{|A|}$.
For simplicity, when the context is clear, we omit the subscript $i$ in $a_i$.
For each aspect $a$, there exists a finite vocabulary of aspect values, represented as $V_a$, along with a corresponding embedding table $E_a \in \mathbb{R}^{|V_a| \times H}$ that contains embeddings for each value of aspect $a$. 
Figure \ref{fig:al_pic} shows the aspect learning in two state-of-the-art multi-aspect dense retrievers \cite{madr}. 
Both approaches utilize content text as the encoder input. 
Specifically, MTBERT reuses \texttt{CLS} to represent aspects, whereas MADRAL constructs embeddings for the $|A|$ aspects by attending to the final layer of content tokens. 
Both methods train the aspect embeddings by predicting the corresponding value annotation ID in $V_a$ for each of the $|A|$ aspects.

\vspace*{-2mm}
\paratitle{Multi-Granularity.} 
Different granularities of text strings capture semantic information at varied levels. Coarse grains such as sentences or phrases often express more specific intent than finer grains like words or tokens.
Therefore, relying solely on phrase-level value prediction, as previous methods do, might not yield effective aspect representations.
For example, if a product category value is "handmade products", its word-level granularity values would be "$ [ handmade, products ] $", and its token-level granularity values would be "$ [ hand, \#\#made, products ] $".
Formally, we denote the set of granularities as $G$, where each $g$ (with $g\in G$) represents a specific granularity.
In this paper, we use three granularities: $G=\{phrase, word, token\}$. 
We use $V_a^g$ to represent the value vocabulary obtained by decomposing aspect $a$'s values at granularity $g$.
The corresponding aspect value embedding table becomes ${E}_a^g \in \mathbb{R}^{|V_a^g| \times H}$. We list the frequently used notations in Table \ref{tab:notations}.

\vspace*{-2mm}
\section{METHODOLOGY}
\label{sec:method}
In this section, we propose a MUlti-granulaRity-aware Aspect Learning model (\baby) for multi-aspect dense retrieval and introduce its core components.
As in \cite{madr}, \baby is also based on BERT \cite{bert}.
Since \baby encodes both items and queries in the same way, we only use items for illustration.

\begin{figure*}
    \centering
    \setlength{\abovecaptionskip}{0pt}
    \setlength{\belowcaptionskip}{-0.4cm}
    \includegraphics[scale=0.28]{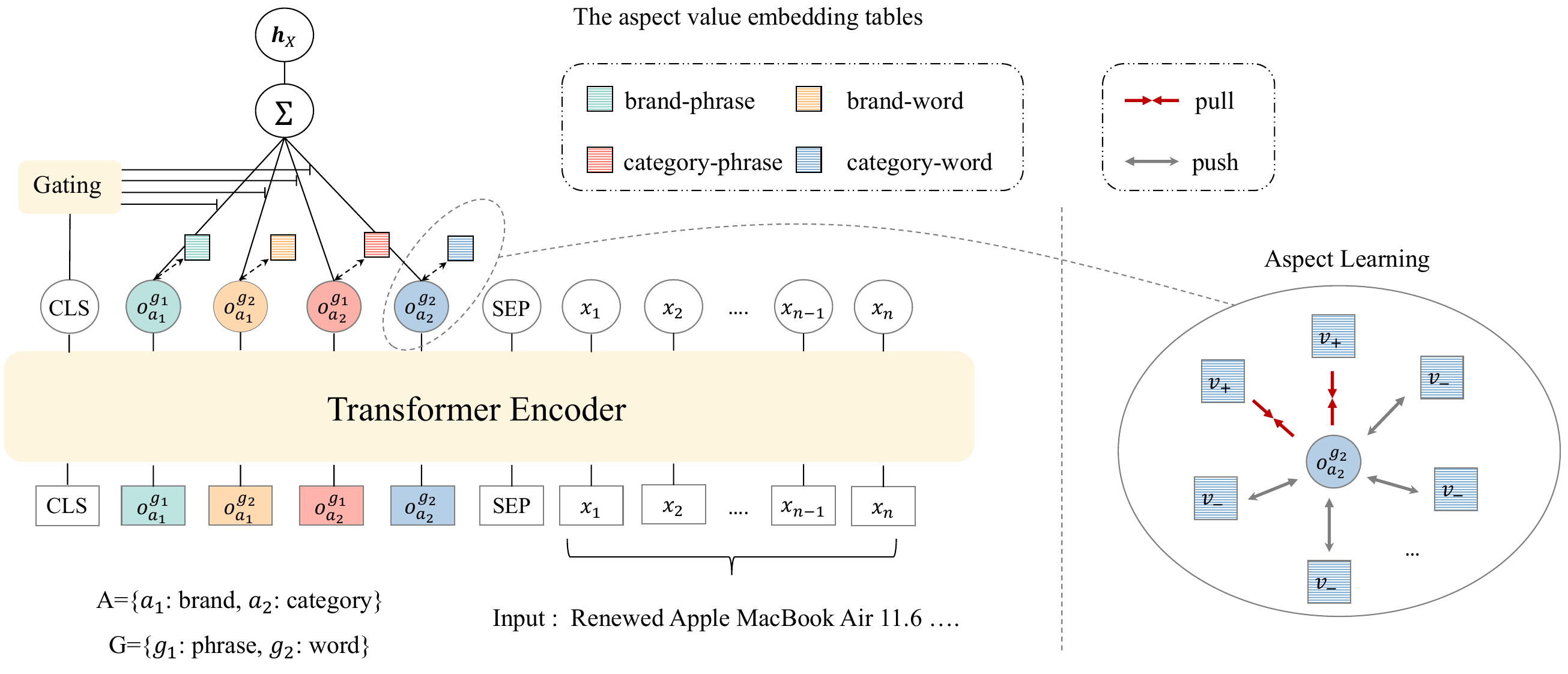}
    \caption{
    Our \baby with single-objective-based grouping in a simplistic scenario of two aspects and two granularities.
    }
    \label{fig:model}
\vspace*{-1mm}
\end{figure*}

\vspace*{-2mm}
\subsection{Aspect Representations}\label{sec:view}
It is crucial to represent aspects reasonably in a pre-trained model so that aspect learning can guide their training effectively. To fully exploit the capabilities of the Transformer encoders, as shown in Figure \ref{fig:model}, we introduce several tokens after \texttt{CLS} and before the content tokens to represent aspects from various perspectives. This aligns with the way \texttt{CLS} is obtained and these tokens can interact with the content tokens sufficiently. During pre-training, these inserted embeddings can act as different views of context when predicting the masked tokens in the content. In this way, these embeddings can also learn from the masked language model objective and capture the content semantics from various implicit views, which could bring more benefits, especially when there are no value annotations for an aspect. 

\vspace*{-2mm}
\paratitle{Comparison with Previous Methods}. 
As shown in Figure \ref{fig:al_pic}, MTBERT\cite{madr} reuses the \texttt{CLS} token to predict the values of item aspects, which enforces \texttt{CLS} to mix the information from item aspects with the overall content semantics it is originally designated to capture. The balance between the two cannot be automatically learned and \texttt{CLS} could be confused about what it should learn. 
MADRAL \cite{madr} represents each item aspect separately by attending to the final representations of content tokens and learns the aspect embeddings by predicting their associated values. During pre-training, the only guidance for the aspect embeddings is this value prediction objective, which could be insufficient to learn them well. What is worse, they will not be updated when there are no aspect-value annotations.
In contrast, in \baby, the aspect information would not mix with the overall content semantics in \texttt{CLS}. With the gating mechanism, they can interact more and the aspect importance can be learned automatically. Moreover, the aspect embeddings can be guided by the masked language model loss as well, which not only benefits their representation learning but also facilitates implicit aspect learning without aspect-value annotations.  

\vspace*{-3mm}
\subsection{Aspect Learning}  \label{sec:al}
For simplicity, we use an example of aspect $a$ at granularity $g$ to illustrate. 
For aspect learning in \baby, there are two important components: value representation and the aspect learning objective. 

\vspace*{-2mm}
\paratitle{Value Representation.} 
To conduct effective aspect learning by predicting the value annotations of an aspect in terms of both coarse and fine grains, value representations play an important role. There are two options: 
1) Sharing the existing token embeddings in the backbone PLM and calculating the value embeddings of word-level and phrase-level grains by a projection function. It can reuse the semantic information carried in the PLM tokens. However, the token embeddings are learned towards the goals of both PLM and aspect learning, which may interfere with each other. 
2) Declaring separate value embedding tables, which is consistent with \cite{madr} and the research before the PLMs era in \cite{rerank-explain}. The extra value embeddings could serve the model to conduct aspect learning better without other interventions. However, if trained from scratch, these new parameters may be difficult to optimize. 
We refer to these two options as ``shared'' and ``unshared'' respectively in terms of whether to share the underlying token embeddings with the existing encoder. We investigate both ways in our experiments (see Section \ref{sec:break}).

Specifically, for the ``shared'' option, we first tokenize each aspect value $v$ in $V_a^g$ using the BERT tokenizer. Then, we extract their embeddings from BERT's embedding tables and use a projector function on the token embeddings to obtain the corresponding value embedding $e_v$ during training.
In this paper, we adopt average pooling as the projection function since it is simple and produces similar results to other methods in our preliminary experiments.

For the ``unshared'' option, each aspect $a$ has a separate embedding table $E_a^g$ for each granularity $g$, storing the embeddings of its values $V_a^g$.
Instead of training these tables from scratch, we initialize the tables using the average token embeddings in the PLM for each value $v$ at granularity $g$ (the same as the initial embedding in the ``shared'' option). 
This gives the new embeddings a decent starting point in the semantic space and has the freedom to better conduct aspect learning, the benefit of which will be shown in Section \ref{sec:break}. 

\vspace*{-2mm}
\paratitle{Aspect Learning Objective.} 
Once we obtain the representation of aspect $a$ at granularity $g$, denoted as $\textbf{h}_a^g(X) \in \mathbb{R}^H$, we adopt the widely-used group-wise contrastive loss to pre-train the encoder.
It aims to bring the source representation closer to instances in its target group while distancing it from representations of other groups \cite{cl}.
\vspace*{-2mm}
\begin{equation}\label{eq:lag}
\small
    \mathcal{L}_a^g(X) = - \frac{1}{|\mathcal{A}_a^g|} \sum_{v^+\in \mathcal{A}_a^g} log \frac{exp(sim(\textbf{h}_a^g(X) , \textbf{e}_{v^{+}}))}{\sum_{v \in V_a^g}{exp(sim(\textbf{h}_a^g(X), \textbf{e}_v))}},
\vspace*{-2mm}
\end{equation}
where $\textbf{e}_v$/$\textbf{e}_{v^+} \in \mathbb{R}^H$ is the aspect value embedding from $E_a^g$, $sim(\cdot)$ is the dot-product function, and $\mathcal{A}_a^g$ is the set of aspect value annotations for aspect $a$ at granularity $g$. 

\vspace*{-2mm}
\subsection{Multi-Granularity-Aspect Grouping} 
\label{sec:mamg}
Assume there are $|A|$ aspects and $|G|$ granularities, our goal is to facilitate aspect learning for each aspect of each granularity, totally |A|*|G| learning objectives.
A straightforward approach is to use a single representation to handle these multiple objectives. However, this method enforces all the information to be compressed together, severely limiting the learning capacity of each objective.
Therefore, we introduce three grouping schemes to integrate multi-granularities and multi-aspects: Single-objective-based Grouping, Granularity-based Grouping, and Aspect-based Grouping.

\vspace*{-2mm}
\paratitle{Single-objective-based Grouping.} 
As shown in Figure \ref{fig:model}, when there are only a few aspects and granularities, we can directly introduce $|A|*|G|$ tokens in the input sequence $X$ 
to capture the item semantics from $|A|*|G|$ views. Each of them accounts for a single objective among the multi-granularity-aspect combinations.
Specifically, we obtain a sequence of $X = (x_0, o_1, ..., o_{|A| * |G|}, x_1, .., x_n)$. 
We utilize the hidden vector $\textbf{h}(o_k)$($k=1, ..., i * j$) from the final layer as the item representation from the perspective of aspect $a_i$ at the granularity $g_j$.
The aspect learning loss function becomes: 
\vspace*{-1mm}
\begin{equation}\label{eq:flatten}
 \small
\begin{aligned}
\mathcal{L_{A}}(X) &= \frac{1}{|A| * |G|} \sum_{i=1}^{|A|} \sum_{j=1}^{|G|} \mathcal L_{a_i}^{g_j}(X).
\end{aligned}
\end{equation}
When $|A| * |G|$ is large, adding a significant number of tokens can adversely affect the semantic representation of the original input. 
Hence, it becomes essential to further group the objectives across various granularities and aspects. 

\vspace*{-2mm}
\paratitle{Granularity-based Grouping.} 
The same granularity indicates the same level of semantic information, and grouping the objectives at the same grain is a reasonable option. 
In this case, $|G|$ tokens will be inserted into the input sequence, yielding $X = (x_0, o_1, ..., o_{|G|}, x_1, x_2, .., x_n)$. Their encoded representations become $\textbf{h}(o_j)$  ($j=1,...,|G|$), representing the item from the perspective of all the aspect information at granularity $g_j$. 
A single aspect embedding accounts for the loss $\mathcal{L}_{a_i}^{g_j}(X)$ of all the aspects $a_i$ ($i=1, ..., |A|$) of granularity $g_j$, i.e., $\textbf{h}_{a_i}^{g_j}$ ($i=1, ..., |A|$) are the same for granularity $g_j$. The aspect learning objective is:
\begin{equation}\label{eq:lg}
\small
\setlength{\abovedisplayskip}{1pt}
\setlength{\belowdisplayskip}{1pt}
\mathcal{L_{A}}(X) = \frac{1}{|G|} \sum_{j=1}^{|G|} \mathcal{L}_{g_j}(X), \textrm{where }\mathcal{L}_{g_j}(X) = \frac{1}{|A|}\sum_{i=1}^{|A|} \mathcal L_{a_i}^{g_j}(X).
\end{equation}

\vspace*{-1mm}
\paratitle{Aspect-based Grouping.} 
An alternative option is to group the objectives across multi-granularity-aspects by aspects so that different aspect information will not mix together and various levels of granularities could benefit each other. 
Here, we introduce $|A|$ guiding tokens before the content tokens: $X = (x_0, o_1, ..., o_{|A|}, x_1, x_2, .., x_n)$. 
The hidden vector $\textbf{h}(o_i)$ ($i=1, ..., |A|$) captures the representations of all granularities for the input item corresponding to aspect $a_i$.
In particular, when calculating the loss $\mathcal{L}_{a_i}^{g_j}(X)$ using equation \ref{eq:lag}, the representation $\textbf{h}_{a_i}^{g_j}$ of aspect $a_i$ remains consistent across different granularities. 
Under this aggregation method, loss $\mathcal{L_{A}}$ can be reformulated as follows:
\begin{equation}\label{eq:la}
 \setlength{\abovedisplayskip}{1pt}
\setlength{\belowdisplayskip}{1pt}
\small
\mathcal{L_{A}}(X) = \frac{1}{|A|} \sum_{i=1}^{|A|} \mathcal{L}_{a_i}(X), \text{where } \mathcal{L}_{a_i}(X) = \frac{1}{|G|}\sum_{j=1}^{|G|} \mathcal L_{a_i}^{g_j}(X).
\end{equation}

Grouping by granularities or aspects reduces the number of guiding tokens, accommodating scenarios with numerous aspects and granularities. 
Their model architectures stay the same as Figure \ref{fig:model}, 
except that aspect learning objectives for the same granularities or aspects are conducted on the shared token.

\subsection{Aspect Embedding Fusion}\label{sec:fusion}
For efficiency concerns, it is necessary to consolidate multiple embeddings into a single one to minimize storage and computation costs. 
Inspired by \cite{madr}, we adopt the "\texttt{CLS}-Gating" fusion mechanism in \baby.
To illustrate the fusion process, we present an example using the single-objective-based grouping approach discussed in Section \ref{sec:mamg}.
Specifically, we pass the \texttt{CLS} embedding through a linear layer and a softmax function to compute the weighting scores for $\textbf{h}(o_1), ..., \textbf{h}(o_{K})$, where $K=|A|*|G|$:
\begin{equation}
\small
\textbf{w} = Softmax(U\textbf{h}(x_0)+b) \in \mathbb{R}^{K},
\end{equation}
where $U \in \mathbb{R}^{K \times H}$ and $b \in \mathbb{R}^{K}$ are trainable parameters.
Then, we utilize the learned weights to fuse multiple embeddings, thereby obtaining the final encoded representation of the input $X$:
\begin{equation}
\small
\setlength{\belowcaptionskip}{-0.3cm}
\setlength{\abovecaptionskip}{0cm}
\begin{aligned}
\textbf{h}_X &=  \sum_{k=1}^{K} w_k \cdot \textbf{h}(o_k). \\
\end{aligned}
\end{equation}

\subsection{Training Objectives}\label{sec:objective}
\vspace*{-2mm}
\paratitle{Pre-training.} 
As discussed in previous work \cite{Ma2021PROPPW}, the Masked Language Model (MLM) \cite{Taylor1953ClozePA} task could help construct good text representation for IR. Therefore, similar to \cite{madr}, we also adopt MLM as one of the pre-training objectives besides aspect learning. 
\begin{equation}
\small
\begin{aligned}
\mathcal{L}_{MLM}(X)=-\sum_{w \in masked(X)}logP(w | X_{\backslash masked(X)}),
\end{aligned}
\end{equation}
where $X$ means the input sentence, $masked(X)$ and $X_{\backslash masked(X)}$ denotes the masked tokens and the remaining tokens from $X$, respectively.

We then pre-train the Transformer encoder using the aspect learning loss jointly with the MLM loss, as follows,
\begin{equation}
\small
\label{eq:total}
\begin{aligned}
\mathcal{L}_{total}(X)=\mathcal{L}_{MLM}(X)+\lambda \mathcal{L_{A}}(X),
\end{aligned}
\end{equation}
where $\lambda$ is the hyperparameter.

\paratitle{Fine-Tuning.} 
We adopt the following in-batch softmax cross entropy loss $\mathcal{L}_{SCE}$ as the learning objective during fine-tuning.
Note that although the aspect learning loss could also be added during fine-tuning, our experimental results show no significant improvements for all the multi-aspect retrievers. Hence, we omit this objective in this paper.
\begin{equation} \label{eq:sce}
\small
\mathcal{L}_{SCE}=- log \frac{exp(sim(\textbf{h}_Q, \textbf{h}_{I^+})) }{exp(sim(\textbf{h}_Q, \textbf{h}_{I^+})) + {\textstyle \sum_{I^-}} {exp(sim(\textbf{h}_Q, \textbf{h}_{I^-}))}}.
\end{equation}

%% file: exp-settings.tex
\begin{table}
\renewcommand{\arraystretch}{0.5}
\setlength\tabcolsep{11pt} 
\setlength{\abovecaptionskip}{0pt}
\setlength{\belowcaptionskip}{0cm}
\caption{
  Aspect-Related Dataset Statistics. It presents the percentage of queries/items with non-empty aspect values in the pre-training corpus and the aspect value vocabulary sizes at various granularities: phrase, word, and token.
}
\label{tab:data}
\begin{tabular}{ccc}
    \toprule
    &  \textit{MA-Amazon} & \textit{\alipay} \\
    \midrule
    \textit{aspect} & \textit{item} & \textit{item / query} \\
    \midrule
    brand  & $94\%$ (5k,6k,5k) & $0.6\%/44\%$ (9k,11k,3k) \\
    \midrule
    color  & $67\%$ (2k,1k,1k) & -- \\
    \midrule
    category  & $87\%$ (8k,5k,5k) & $90\%/91\%$ (457,650,548) \\
    \bottomrule
\end{tabular}
\end{table}

\begin{table*}[th]
\setlength\tabcolsep{9pt}
\renewcommand{\arraystretch}{0.7}
\setlength{\abovecaptionskip}{0pt}
\setlength{\belowcaptionskip}{-0.3cm}
\caption{Comparisons between \baby and the baselines. 
The best results (excluding \baby-CONCAT) are in bold. 
$\dag$, $\ddag$, and $\ast$ indicate significant improvements over the best baselines in the first/second group and the backbone BIBERT, respectively.
}
  \label{main_exp}
\begin{threeparttable}
  \begin{tabular}{lllllll}
    \toprule
     \multirow{2}{*}{Method} & \multicolumn{3}{c}{MA-Amazon}&\multicolumn{3}{c}{\alipay} \\
     \cmidrule(lr){2-4} \cmidrule(lr){5-7}
      &R@100&R@500 & NDCG@50 & R@100 & R@500  & NDCG@50 \\
    \midrule
    BIBERT& 0.6075 & 0.7795 & 0.3929 & 0.4464 & 0.6284 & 0.2033 \\
    Condenser& \underline{0.6091}&  \underline{0.7801} & \underline{0.3960}& \underline{0.4520} & \underline{0.6423} & \underline{0.2072} \\
    \midrule
    BIBERT-CONCAT& 0.6137 & 0.7814 & \underline{0.4005} &  \underline{0.4517} & 0.6291 & \underline{0.2103} \\
    MTBERT& \underline{0.6137} & \underline{0.7852} & 0.3969 & 0.4498 & 0.6280 & 0.2064   \\
    MADRAL & 0.6088 & 0.7815 &  0.3950 & 0.4506 & \underline{0.6383} & 0.2057   \\
    \midrule
    \babyw & 0.6282$^{\dag \ddag \ast}$ & 0.7943$^{\dag \ddag \ast}$  & 0.4151$^{\dag \ddag \ast}$  & {0.4556}$^{\ast}$ & {0.6458}$^{\ddag \ast}$ & {0.2046}   \\
    \baby& \textbf{0.6371}$^{\dag \ddag \ast}$ & \textbf{0.8023}$^{\dag \ddag \ast}$ &  \textbf{0.4228}$^{\dag \ddag \ast}$ & \textbf{0.4630}$^{\dag \ddag \ast}$ & \textbf{0.6519}$^{\dag \ddag \ast}$ & \textbf{0.2177}$^{\dag \ddag \ast}$   \\
    \midrule
    \baby-CONCAT& {0.6389}$^{\dag \ddag \ast}$ & {0.8005}$^{\dag \ddag \ast}$ &  {0.4281}$^{\dag \ddag \ast}$ & {0.4669}$^{\dag \ddag \ast}$ & {0.6474}$^{\ddag \ast}$ & {0.2124}$^{\dag \ast}$\\
    \bottomrule
  \end{tabular}
  \end{threeparttable}
\vspace*{-3mm}
\end{table*}

\section{EXPERIMENTAL SETTINGS}
\subsection{Datasets}
We use the following two large-scale search datasets from real-world platforms with rich aspect information for our experiments. The aspect-related statistics of the two datasets are in Table \ref{tab:data}.
(1) \textbf{Multi-Aspect Amazon ESCI Dataset (MA-Amazon).} 
MA-Amazon \cite{sun2023cikm} enriches the English portion of the Amazon ESCI \cite{amazon-data} dataset with item category information. In MA-Amazon, only items have annotations for \textit{brand}, \textit{color} and \textit{category}. 
The item corpus contains 482K distinct products, which are used for pre-training. 
The retrieval dataset has 17k, 3.5k, and 8.9k queries for fine-tuning, validation, and testing respectively, without any query overlaps. 
For each query, the retrieval dataset provides 20.1 items on average, along with their ESCI relevance judgments (\textit{Exact}, \textit{Substitute}, \textit{Complement}, \textit{Irrelevant}), indicating each item's relevance to the given query.
Following \cite{amazon-data}, we treat \textit{Exact} as positive and all others as negatives for fine-tuning and metrics requiring binary labels.
(2) \textbf{\alipay Search Dataset.} \alipay is a Chinese mini-program (app-like service) search dataset. 
In \alipay, both queries and items are annotated with two aspects: \textit{brand} and \textit{category}. 
We conduct pre-training on both a query corpus, containing 1.3M unique queries, and an item corpus with 1.8M distinct items.
The retrieval dataset contains 60k, 3.3k, and 3.3k real user queries for fine-tuning, validation, and testing respectively, without query overlaps. Note that the queries for validation and testing do not appear in the pre-training query corpus. Each instance in the relevance dataset is a <\textit{query}, \textit{item}, \textit{label}> triplet, where the label indicates the manually annotated binary relevance of this query-item pair. 


\vspace*{-2mm}
\subsection{Baselines}
We adopt the following dense retrieval baselines for comparison, including models using or without using aspect information:
(1) \textbf{BIBERT} \cite{bibert-pretrain,sent-bert}: A standard bi-encoder baseline and the backbone of \baby, using \texttt{CLS} encoding of the BERT-based encoder for both query and item representations. BIBERT is pre-trained with MLM loss and fine-tuned with loss $\mathcal{L}_{SCE}$ (Equation \ref{eq:sce}).
(2) \textbf{Condenser}\cite{condenser}: 
A pre-trained method tailored for unstructured textual dense retrieval.
It introduces a short circuit between middle-layer tokens (excluding \texttt{CLS}) and their corresponding head-layer tokens during pre-training, optimizing the \texttt{CLS} embedding to encapsulate more information.
(3) \textbf{BIBERT-CONCAT}: It treats the aspect values as texts and concatenates them with the query/item content during pre-training with MLM. During fine-tuning, since the concatenation with query could change query semantics 
, we only concatenate item aspects for relevance matching. 
(4) 
\textbf{MTBERT}\cite{madr}: A multi-task (MT) learning model based on BIBERT. Besides MLM during pre-training, it conducts $|A|$ aspect prediction tasks using \texttt{CLS}.
(5)
\textbf{MADRAL}\cite{madr}: 
It incorporates an aspect extraction attention network to extract $|A|$ aspect representations for both queries and items. These embeddings are learned from aspect prediction tasks during pre-training and fused to yield the final representation during fine-tuning. 
(6) \textbf{\baby, \babyw and \baby-CONCAT}: 
\baby is our proposed multi-aspect dense retrieval model. 
In contrast, \babyw disables aspect learning. Specifically, when $\lambda$ in Equation \ref{eq:total} is set to 0, \baby regresses to \babyw.
\baby-CONCAT employs the same aspect-content text concatenation strategy as BIBERT-CONCAT for the model input.
Note that unless the model name includes "-CONCAT", the model input consists solely of content text.


\subsection{Implementation Details}
We implemented \baby and all the baselines by ourselves to ensure consistent implementation details and fair comparisons. 

\subsubsection{Pre-training}
For all methods, the encoder is shared for both queries and items to facilitate knowledge sharing. 
Specifically, we pre-train on a corpus consisting of the item corpus or a mixture of the query and item corpus (when query aspect annotations are available) to obtain the shared encoder for fine-tuning.

\noindent%
\textbf{Muti-granularity Value Collection.} Given an aspect $a$ and its original aspect vocabulary at the phrase level, we obtain its word and token granularity vocabularies: For word granularity, we segment each aspect value $v$ by spaces and punctuation (for English) or employ the Jieba tool \cite{jieba} (for Chinese), and eliminate duplicates to aggregate the generated ``words''.
For token granularity, we merge the token list obtained by processing each aspect value $v$ with the BERT tokenizer to create the corresponding vocabulary set.

\noindent%
\textbf{Model Pre-training.}
We initialize all the BERT components using Google's public checkpoint and employ the Adam optimizer with the linear warm-up technique. 
The learning rate and epoch for the MA-Amazon/\alipay dataset are set to 1e-4/5e-5 and 20/10, respectively. The maximum token length is 156, the MLM mask ratios are 0.15 for items and 0.3 for queries.
For all methods requiring adjustment of training objective scaling coefficients, we uniformly select coefficients based on their validation set performance after fine-tuning. 
These coefficients vary from 0.1 to 1, in 0.1 intervals. 
For our method, we set $\lambda$ in Eq.\ref{eq:total} to 0.1. 
 We use the following fine-tuning procedures to evaluate pre-trained model checkpoints every two epochs and select the best one on the validation set.

\subsubsection{Fine-tuning}
For both datasets, we fine-tune all the models for 20 epochs with Tevatron toolkit\cite{tevatron}. Following the previous work \cite{DPR}, we include a hard negative sample for each query besides in-batch negatives. We use a learning rate of 5e-6 and a batch size of 64. The maximum token lengths are set at 32 for queries and 156 for items. All the models are trained with relevance loss $\mathcal{L}_{SCE}$ (Eq.\ref{eq:sce}). 

\subsubsection{Evaluation Metrics}
We report R@100, R@500, and NDCG@50. Following \cite{amazon-data}, we assign the gains of 1.0, 0.1, 0.01, and 0.0 to E, S, C, and I, respectively, for MA-Amazon. We conduct two-tailed paired t-tests (p < 0.05) to identify significant differences.

%% file: exp-results.tex
\vspace*{-2mm}
\section{EXPERIMENT RESULTS}

\subsection{Overall Performance}
\label{exp:overall}
We compare \baby with baseline models both utilizing and without utilizing aspect information.
As shown in Table \ref{main_exp}, we have the following observations: 
(1) The models that leverage the aspect information (methods except for BIBERT, Condenser, and MUR) outperform their backbone (BERT) without using it. Among the multi-aspect dense retrievers, \baby performs the best with a significantly large margin.  
This confirms the necessity of incorporating aspects in query/item representation learning.
(2) 
On MA-Amazon, MADRAL underperforms the simpler MTBERT. We believe this is due to the less pre-training data of MA-Amazon and the absence of aspect annotations for queries, which makes the aspect embeddings of MADRAL not sufficiently learned. In contrast, \baby achieves compelling performance consistently.
(3) 
Condenser, a more advanced pre-trained model for unstructured text retrieval, sometimes outperforms the baseline multi-aspect dense retrievers. Notably, the gains from advanced pre-training techniques are orthogonal to our method. Our approach can be easily incorporated into stronger backbones like Condenser and could achieve even better performance. We leave it in our future work.
(4)
BIBERT-CONCAT performs better than MTBERT and MADRAL in terms of some metrics on the two datasets, indicating that concatenating aspects as text strings can be beneficial. However, query aspects should be taken special care of during relevance matching in order to achieve good performance. 
Incorporating both concatenation and aspect prediction in the same model (\baby-CONCAT) does not always result in better performance than without concatenation. 
We have similar observations with MTBERT and MADRAL, but due to the space concern, we do not report them. 
The reason may be that the model learns unwanted shortcuts when using the aspect both as the model input and the learning objective.
(5) 
Our method shows competitive performance even without using aspect annotations (\babyw). \babyw outperforms most baseline models in terms of all the metrics except NDCG@50 on \alipay. This indicates that \babyw can capture complementary information from implicit perspectives for the final representation. It also confirms the advantages of aspect representations and the MLM training for the aspects in \baby. 

\vspace*{-2mm}
\subsection{Studies on Model Variants}\label{sec:break}
In this subsection, we study various options for the essential components in \baby. 
For reproducibility, all experiments are conducted on the public MA-Amazon dataset.

\begin{table}
\setlength{\abovecaptionskip}{0pt}
    \caption{Variants of \baby on MA-Amazon. 
    $\dag$ indicates significant differences from the best option.}

\renewcommand{\arraystretch}{0.6}
    \setlength\tabcolsep{6.8pt}
    \centering
    \label{tab:option2}
    \begin{tabular}{llllllll}
    \toprule
        Method & R@100 & R@500  & NDCG@50 \\
        \midrule
        {\baby$^{unshared\_single}$}  & 0.6336\rlap{$^{\dag}$} & 0.8005  & 0.4195\rlap{$^{\dag}$} \\
        {\textbf{\baby}$^{unshared\_granu}$}  & \textbf{0.6371} & \textbf{0.8023}  & \textbf{0.4228} \\
        {\baby$^{unshared\_aspect}$} & 0.6340\rlap{$^{\dag}$}  & 0.8003\rlap{$^{\dag}$}  & 0.4195\rlap{$^{\dag}$} \\
        \midrule
        {\baby$^{shared\_single}$}  & 0.6300\rlap{$^{\dag}$} & 0.7980\rlap{$^{\dag}$}  & 0.4171\rlap{$^{\dag}$} \\
        {\baby$^{shared\_granu}$}  & 0.6333\rlap{$^{\dag}$} &  0.7996\rlap{$^{\dag}$} &  0.4184\rlap{$^{\dag}$} \\
        {\baby$^{shared\_aspect}$}  & 0.6321\rlap{$^{\dag}$} & 0.7995\rlap{$^{\dag}$}  & 0.4173\rlap{$^{\dag}$} \\
        \midrule
        {\baby$^{unshared\_randinit}$} & 0.6337\rlap{$^{\dag}$}  & 0.8006 & 0.4196\rlap{$^{\dag}$} \\
        \midrule
        \midrule
        {\baby$^{first\_k}$} & 0.6141\rlap{$^{\dag}$} & 0.7873\rlap{$^{\dag}$} & 0.4009\rlap{$^{\dag}$} \\
        \midrule
        \midrule
        {\baby$^{no\_cls\_gating}$} & 0.6212 \rlap{$^{\dag}$} & 0.7890 \rlap{$^{\dag}$} & 0.4064\rlap{$^{\dag}$} \\
    \bottomrule
    \end{tabular}
\end{table}

\paratitle{Studies on Value Representations.} 
In Table \ref{tab:option2}, we observe that using an independent value embedding space (the ``unshared'' option in Section \ref{sec:al}) leads to better performance. As we mentioned earlier, the ``shared'' option optimizes token embeddings both towards the objectives in BERT and the aspect learning, which could interfere with each other and limit the capacity of the model on aspect prediction.  
However, under the ``unshared'' option, it may be difficult to optimize the separate value embeddings from scratch while other parameters only need fine-tuning. To see whether this affects model performance, instead of using the same initial state as the ``shared'' option, we randomly initialize the embeddings while keeping other best settings in \baby. The harmed performance of \baby$^{unshared\_randinit}$ in Table \ref{tab:option2} confirms our presumption and shows the benefit of decent initialization states.   

\paratitle{Studies on Grouping Methods.}
In Table \ref{tab:option2}, we observe that \baby$^{unshared\_granu}$, which groups objectives by granularity, performs the best. 
Note that on the \alipay dataset, which has fewer aspects, \baby$^{unshared\_single}$, single-objective-based grouping, has the best performance. 
This is consistent with our claim in Section \ref{sec:mamg} that as the aspect count increases, further grouping benefits model training.
Based on these observations, we suggest: 
(1) For small numbers of aspects and granularities, simply use independent learning for each objective (\baby$^{unshared\_single}$).
(2) When there are more aspects and granularities, grouping multiple objectives in one guiding token can be a better choice.  
 

\paratitle{Studies on Guiding Tokens and Fusion Methods.}
Instead of adding separate guiding tokens for aspect learning, we study a variant that reuses the same amount of tokens at the beginning of the input sequence to conduct aspect learning, denoted as \baby$^{first\_k}$. The results show that \baby$^{first\_k}$ has similar or better performance to the best baseline in Table \ref{main_exp} but is significantly worse than the best variant of \baby. This indicates that the multi-granularity-aware aspect learning is beneficial but using separate guiding tokens to conduct the learning is needed. 

To study whether \texttt{CLS}-Gating (introduced in Section \ref{sec:fusion}) is helpful for aspect embedding fusion, in \baby$^{no\_cls\_gating}$, 
we remove it and use the \texttt{CLS} embedding as the final representation. \texttt{CLS} naturally fuses the aspect embeddings in the second-to-last layer while the aspect learning is conducted in the last layer.  
This variant performs better than the best baseline in Table \ref{main_exp} but is worse than the best variant. It indicates that the fusion should be carried on the final aspect embeddings with a proper weighting mechanism.

\begin{table}
\setlength{\abovecaptionskip}{0cm}
\setlength{\belowcaptionskip}{-0.1cm}
    \caption{Ablation studies of \baby in terms of category and granularity on the MA-Amazon dataset. $\dag$, $\ddag$ indicate significant differences over \baby and BIBERT.}
\renewcommand{\arraystretch}{0.7}
    \setlength\tabcolsep{7.8pt}
    \label{tab:ab-aspect}
    \centering
    \begin{tabular}{lllll}
    \toprule
        Method & R@100 & R@500  & NDCG@50 \\
    \midrule
    BIBERT & 0.6075 & 0.7795 & 0.3929 \\
    \baby & \textbf{0.6371} & \textbf{0.8023} &  \textbf{0.4228} \\
    \midrule
        \baby$^{\  only \  brand}$ & 0.6289\rlap{$^{\dag \ddag}$} & 0.7951\rlap{$^{\dag \ddag}$} & 0.4168\rlap{$^{\dag \ddag}$} \\
        \baby$^{\  only \  color}$ & 0.6284\rlap{$^{\dag \ddag}$} & 0.7942\rlap{$^{\dag \ddag}$} & 0.4158\rlap{$^{\dag \ddag}$} \\
        \baby$^{\  only \  category}$ & 0.6315\rlap{$^{\dag \ddag}$} & 0.7994\rlap{$^{\dag \ddag}$} & 0.4166\rlap{$^{\dag \ddag}$} \\
    \midrule
        \baby$^{\  only \  phrase}$ & 0.6315\rlap{$^{\dag \ddag}$} & 0.7983\rlap{$^{\dag \ddag}$} & 0.4185\rlap{$^{\dag \ddag}$} \\
        \baby$^{\  only \  word}$ & 0.6309\rlap{$^{\dag \ddag}$} & 0.7994\rlap{$^{\dag \ddag}$} & 0.4194\rlap{$^{\dag \ddag}$} \\
        \baby$^{\  only \  token}$ &  0.6305\rlap{$^{\dag \ddag}$} & 0.7982\rlap{$^{\dag \ddag}$} & 0.4192\rlap{$^{\dag \ddag}$}  \\
    \bottomrule
    \end{tabular}
\end{table}

\vspace*{-2mm}
\subsection{Ablation Studies}\label{sec:ablation}
We ablate various components of multi-aspect, multi-granularity, and query/item aspect learning.
In this section, our experiments are also based on the enriched MA-Amazon dataset. 
Additionally, we validate the importance of the query and item side effects on the \alipay dataset, since MA-Amazon lacks query-side aspect information. 
\vspace*{-2mm}
\begin{table}
\vspace*{-1mm}
\setlength{\abovecaptionskip}{0pt}
    \caption{Ablation of query and item aspects on \alipay.  
    $\dag$, $\ddag$ indicate significant differences over \baby and BIBERT.
    }
    \label{tb:qd}
\renewcommand{\arraystretch}{0.7}
    \setlength\tabcolsep{10.7pt}
    \centering
    \begin{tabular}{llll}
    \toprule
        Method & R@100 & R@500  & NDCG@50 \\
    \midrule
    BIBERT & 0.4464 & 0.6284 & 0.2033 \\
    \baby & \textbf{0.4630} & \textbf{0.6519} & \textbf{0.2177} \\
    \midrule
    \baby$^{-query}$ & 0.4569\rlap{$^{\ddag}$} & 0.6400\rlap{$^{\dag \ddag}$} & 0.2126\rlap{$^{\dag \ddag}$} \\
    \baby$^{-doc}$ & 0.4573\rlap{$^{\ddag}$} & 0.6454\rlap{$^{\dag \ddag}$} & 0.2103\rlap{$^{\dag \ddag}$} \\
    \bottomrule
    \end{tabular}
\end{table}

\vspace*{-2mm}
\paratitle{Effect of Aspects and Granularities.}
In Table \ref{tab:ab-aspect}, we first study the impact of multi-aspect and multi-granularity in \baby. 
We find that: 
(1) Every aspect contributes to the model performance, especially \textit{category}, consistent with \cite{madr}. 
(2) Each granularity alone is conducive to model performance and combining them all leads to even better results. Different granularities capture semantics at distinct levels and they can complement each other. 

\vspace*{-2mm}
\paratitle{Effect of Query/Item Aspects.}
We disable the aspect learning on the query/item side in Table \ref{tb:qd}. 
We observe that both query and item aspects are beneficial
and query aspects have a larger impact, which is also consistent with \cite{madr}. 
This is not surprising since query aspects are obtained from query analysis such as intent classification and carry more additional information. 


\begin{table}
\vspace*{-2mm}
\setlength{\abovecaptionskip}{0pt}
    \caption{The category aspect accuracy on \alipay dataset. }
    \label{tb:acc}
\renewcommand{\arraystretch}{0.6}
    \setlength\tabcolsep{4pt}
    \centering
    \begin{tabular}{c|cccccccccc}
    \toprule
         \multirow{2}{*}{Method} & & \multicolumn{2}{c}{query} & \multicolumn{2}{c}{doc} \\
         \cmidrule(lr){3-4} \cmidrule(lr){5-6} 
         & & pre-train & fine-tune & pre-train & fine-tune   \\
         \midrule
         {MTBERT} & phrase & 0.86 & 0.11 & 0.97 & 0.20	 \\
         \midrule
         {MADRAL} & phrase & 0.88  & 0.81 & 0.98 & 0.87  \\
         \midrule
         \multirow{3}{*}{\baby} & phrase & 0.89 & 0.85 & 0.98 & 0.93  \\
         & word & 0.88 & 0.82  & 0.97 &  0.93  \\
          & token & 0.89  & 0.73 & 0.97 & 0.80  \\
    \bottomrule
    \end{tabular}
\end{table}

\vspace*{-2mm}
\subsection{Aspect Learning Accuracy}
In Table \ref{tb:acc}, we compare the accuracy of \baby with baseline methods after pre-training and fine-tuning to understand the aspect learning process better. 
We only analyze the most important aspect - \textit{category} on \alipay. MA-Amazon and other aspects have similar conclusions. 
Considering that each item may have multiple category annotations, we use Accuracy@3 to calculate accuracy.
Evaluation of the query and item aspect prediction is based on the test query set and item corpus of the \alipay dataset, respectively. 

First, all methods have high accuracy after aspect learning in pre-training while lower accuracy after fine-tuning. 
Since we only use relevance loss during fine-tuning, it is expected that the accuracy will drop. 
In our experiments, we find that adding aspect learning loss during fine-tuning enhances aspect prediction accuracy but will harm retrieval performance. We speculate that this objective guides the model parameters to somewhere not aligned with the relevance-matching objective. Hence, higher aspect prediction accuracy does not always co-occur with better retrieval performance. 

Secondly, the prediction accuracy of MTBERT drops dramatically after fine-tuning. Since MTBERT uses the same \texttt{CLS} token to conduct relevance matching and aspect prediction, optimization only toward relevance matching during fine-tuning undermines its ability to predict aspect values. 
In contrast, MADRAL and \baby retain most of such ability after fine-tuning since they use extra aspect embeddings to perform aspect learning.  

Lastly, for phrase-level evaluation, \baby has the best aspect prediction accuracy. As we know, \baby also has the best retrieval performance, which means \baby can learn the two objectives well and let better aspect embeddings assist relevance matching more. 
Notably, the accuracy at the word and token level is not comparable with the phrase level since the ground truth is different. 
The finer-level prediction accuracy is also good. When the granularity becomes finer, the accuracy becomes lower after fine-tuning, which is probably because finer grains have more ground-truth values, making the multi-label classification more challenging.

\begin{figure}
\setlength{\belowcaptionskip}{-0.1cm}
\setlength{\abovecaptionskip}{0cm}
  \centering
  \includegraphics[scale=0.32]{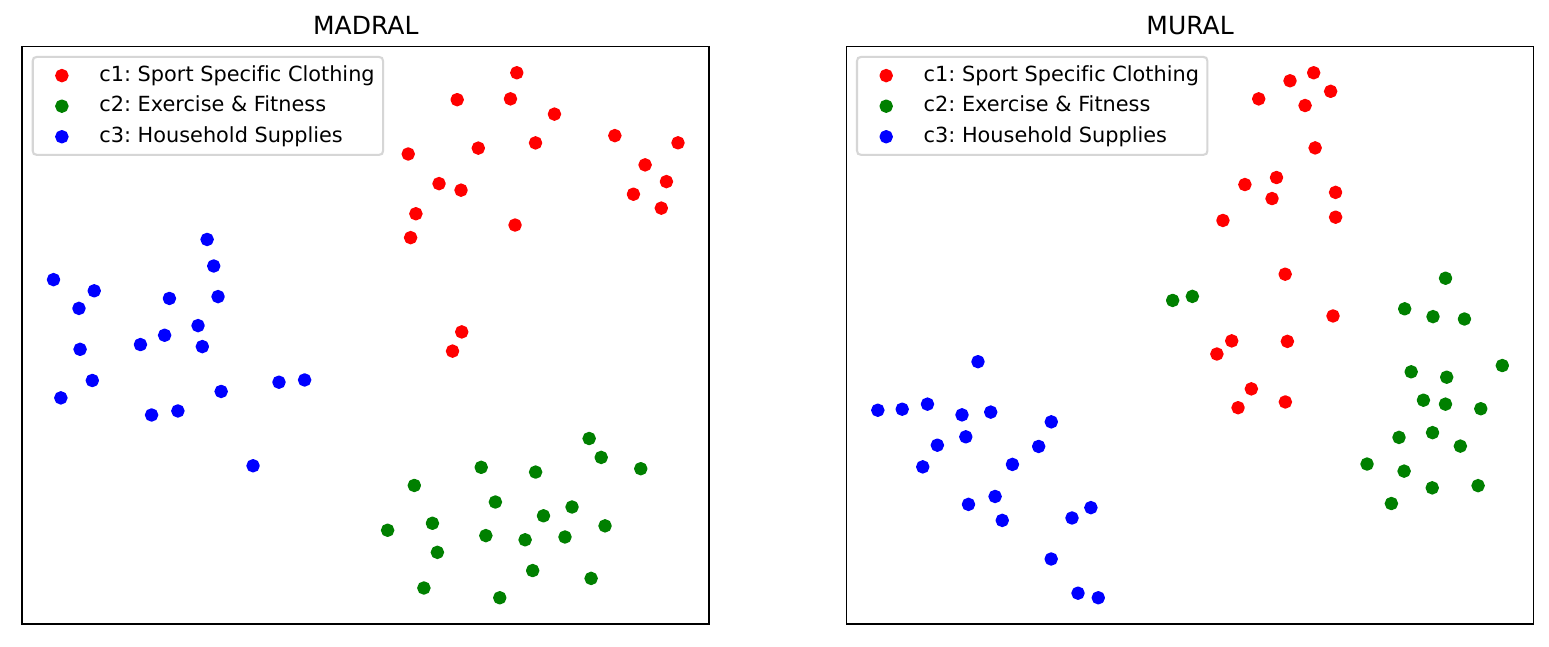}
  \caption{The t-SNE plot of item representations for MADRAL and \baby on MA-Amazon.}
  \label{fig:tsne}
\end{figure}
\subsection{Case Visualization}
We visualize the item representations of three categories, as shown in Figure \ref{fig:example}, to see their distributions in semantic space. Specifically, \textit{c1} (Sport Specific Clothing) and \textit{c2} (Exercise \& Fitness) are semantically similar, while \textit{c3} (Household Supplies) is unrelated to the first two.
We first use MADRAL and \baby to obtain all the item representations on MA-Amazon and put items into their categories. 
Then we randomly pick 20 items of \textit{c1}, \textit{c2} and \textit{c3} and plot them using the t-SNE toolkit in Figure \ref{fig:tsne}. We can observe that MADRAL separates \textit{c1}, \textit{c2}, and \textit{c3} to a similar extent. By contrast,
\baby places the related categories \textit{c1} and \textit{c2} closer and puts them farther from the unrelated \textit{c3}.  
This demonstrates MADRAL's inability to discern the semantic similarity between \textit{c1} and \textit{c2}, as it treats different phrase-level product categories as isolated IDs, overlooking their word-level semantic connections. 
\baby can capture fine-grained semantic relations among similar aspect values while maintaining precise phrase-level aspect discrimination by incorporating both coarse and fine granularity information. 